\documentclass[prb,amsmath,amssymb,floatfix,superscriptaddress,twocolumn]{revtex4}
\usepackage{bm}
\usepackage{url}
\usepackage{graphicx}
\usepackage{lipsum}
\usepackage{physics}

\usepackage{epsfig}
\usepackage{subfigure}
\usepackage[usenames]{color}
\usepackage{hyperref}

\newcommand{\anp}[1]{{\color{blue}#1}}

\let\anp\relax
\usepackage[normalem]{ulem}
\newcounter{para}

\newcommand{\beginsupplement}{%
        \setcounter{table}{0}
        \renewcommand{\thetable}{S\arabic{table}}%
        \setcounter{figure}{0}
        \renewcommand{\thefigure}{S\arabic{figure}}%
     }

\begin{document}

\title{High-Throughput {\it Ab Initio} Design of Atomic Interfaces using InterMatch}

\author{Eli Gerber}
\affiliation{School of Applied and Engineering Physics, Cornell University, Ithaca, NY 14853, USA}

\author{Steven B. Torrisi}
\affiliation{Department of Physics, Harvard University, Cambridge, MA 02138, USA}
\affiliation{Energy \& Materials Division, Toyota Research Institute, Los Altos, CA 94022, USA}

\author{Sara Shabani}
\affiliation{Department of Physics, Columbia University, New York, NY, USA}

\author{Eric Seewald}
\affiliation{Department of Physics, Columbia University, New York, NY, USA}

\author{Jordan Pack}
\affiliation{Department of Physics, Columbia University, New York, NY, USA}

\author{Jennifer E. Hoffman}
\affiliation{Department of Physics, Harvard University, Cambridge, MA 02138, USA}
\affiliation{John A. Paulson School of Engineering and Applied Sciences,Harvard University, Cambridge, MA 02138, USA}

\author{Cory R. Dean}
\affiliation{Department of Physics, Columbia University, New York, NY, USA}

\author{Abhay N. Pasupathy}
\affiliation{Department of Physics, Columbia University, New York, NY, USA}

\author{Eun-Ah Kim}
\affiliation{Department of Physics, Cornell University, Ithaca, NY 14853, USA}

\begin{abstract}
Forming a hetero-interface is a materials-design strategy that can access an astronomically large phase space. However, the immense phase space necessitates a high-throughput approach for an optimal interface design. Here we introduce 
 a high-throughput computational framework, InterMatch, for efficiently predicting charge transfer, strain, and superlattice structure of an interface by 
 leveraging the databases of individual bulk materials. Specifically, the algorithm reads in the lattice vectors, density of states, and the stiffness tensors for each material in their isolated form from the Materials Project. From these bulk properties, InterMatch estimates the interfacial properties. We benchmark InterMatch predictions for the charge transfer against experimental measurements and supercell density-functional theory calculations. We then use InterMatch to predict promising interface candidates for doping transition metal dichalcogenide MoSe$_2$. Finally, we explain experimental observation of factor of 10 variation in the supercell periodicity within a few microns in graphene/$\alpha$-RuCl$_3$ by exploring low energy superlattice structures as a function of twist angle using InterMatch. We anticipate our open-source InterMatch algorithm accelerating and guiding ever-growing interfacial design efforts.
 Moreover, the interface database resulting from the InterMatch searches presented in this paper can be readily accessed through \href{https://contribs.materialsproject.org/projects/intermatch/}{MPContribs}. 

\end{abstract}

\date{\today \ [file: \jobname]}
 \maketitle

\begin{figure*}[t]
\begin{centering}
\includegraphics[width=.9\textwidth]{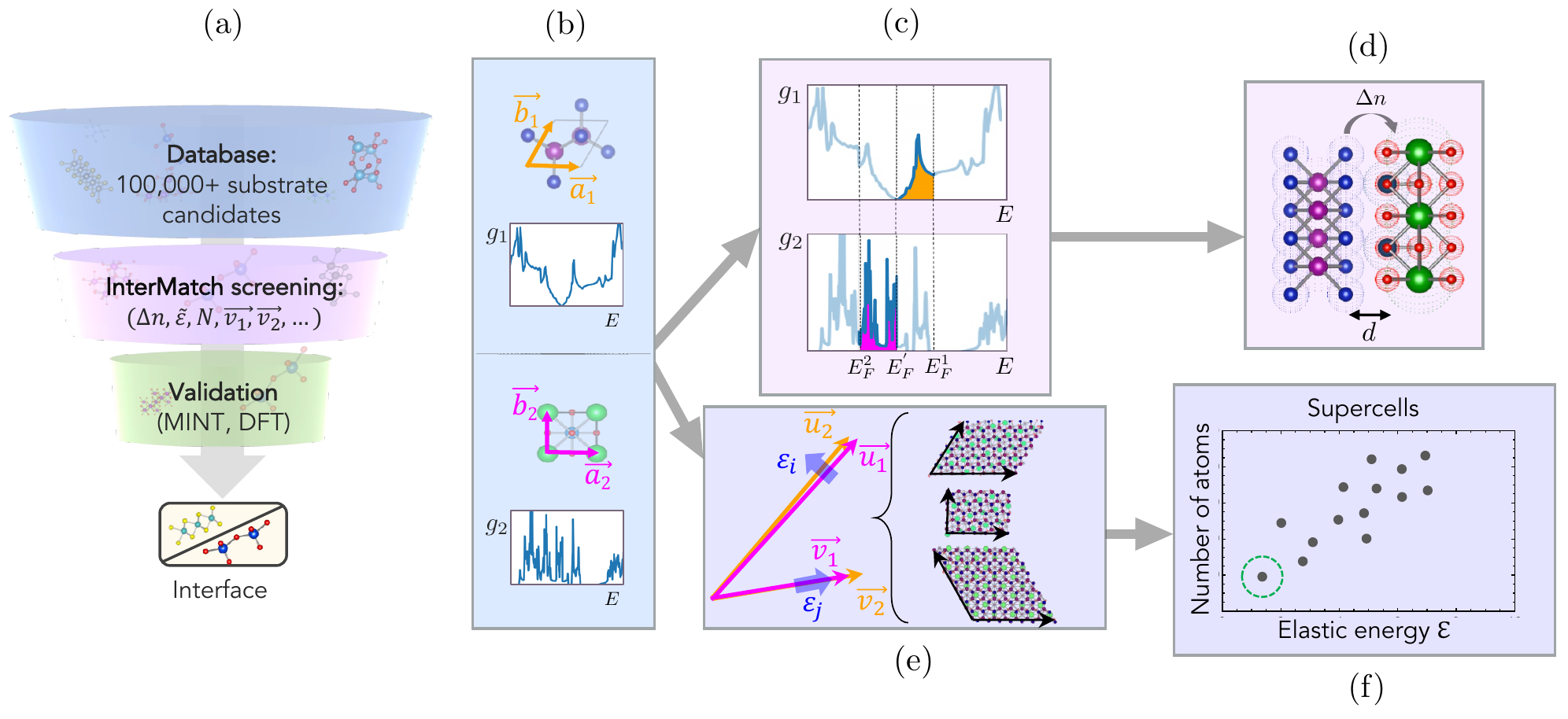}
\end{centering}
\caption{(a) Role of InterMatch in the materials discovery process. (b) Input from the bulk database. Lattice vectors $\vec{a}_\alpha$, $\vec{b}_\alpha$, density-of-states $g_\alpha$ of systems $\alpha=1,2$. Elastic tensors are additional inputs. (c) $E_F^{\alpha}$ are bulk Fermi levels and $E_F'$ is new equilibrium Fermi level.  
(d) $\Delta n$ is the transferred charge density and $d$ is the interlayer separation between the two systems, taken to be the sum of the largest van der Waals radii of the species in each system 1 and 2.
 (e) Superlattice vectors $\vec{v_i}$ (orange arrows) and their near-equivalent vectors $\vec{u_i}$ (magenta arrows). 
Candidate supercells, formed in each basis by combining $\{(\vec{v}_i,\vec{u}_i),(\vec{v}_j,\vec{u}_j)\}$ pairs specify the strain $\varepsilon_i$ (blue arrows). 
(f) Optimal supercell minimizes the elastic energy and the number of atoms in the cell. 
}
\label{Fig:fig1}
\end{figure*}

With increasing control in interface fabrication, interfacial systems form an arena of limitless possibilities\cite{Geim2013Nature}. Recent developments with moir\'e heterostructures\cite{Kennes2021Nat.Phys.} further enlarged the phase space to include the twist angle. However, the vast space of possibilities also implies it is crucial to go beyond serendipitous discoveries and empirical explorations to effectively harness the intrinsic potential of interfacial systems. The traditional approach to theoretically studying interfaces is to carry out density-functional theory (DFT) calculations on a supercell system consisting of two materials\cite{Komsa2013PRB,Terrones2014JMR,Ebnonnasir2014APL,Bokdam2014PRB,Zhou2015Phys.Rev.B}. While such approaches are rigorous, computational limitations regularly require imposing unnatural strain to form a periodic structure.
Moreover, the $\mathcal{O}(N^{3})$ scaling of DFT in the number of electrons $N$ for each such calculation prohibits a comprehensive exploration. Some of us recently proposed an ``intermediate scale'' approach called Mismatched INterface Theory (MINT)\cite{Gerber2020Phys.Rev.Lett.}, which can predict charge transfer and natural strain approximating one layer of the interface using finite-size scaling of atomic clusters. While MINT calculations are much more computationally affordable, they are not fast enough for an exhaustive survey in real-time.  
Therefore, a comprehensive and fast approach to scanning the relevant phase space of interfacial combinations is greatly needed. 

With advancements in widely available comprehensive materials databases\cite{Horton2021NatComputSci,Levin2020,Landis2012IEEE,CURTAROLO2012227,Jain2013APLMaterials,Wolverton2015JOM,Asta2015Sci.Data,Ashton2017Phys.Rev.Lett.,Haastrup20182DMater.,Zhou2019Sci.Data,Draxl2019JMP,Talirz2020Sci.Data,Choudhary2020npjComp.Mat.}, it is timely to establish a high-throughput approach to interface design that can leverage the information contained in these databases to make predictions of interface physics. While it is possible to look up a pair of materials in the database and try to match their properties, the phase space of all possible combinations amounts easily to $>\mathcal{O}(10^{6})$ possibilities, which is beyond the scale of what a manual search can reasonably accomplish. At the same time, it is desirable to benefit from extensive existing databases and search exhaustively for ideal combinations to reach a given objective. Indeed, efforts to make interfacial predictions using bulk material databases are beginning to emerge\cite{Mathew2016ComputationalMaterialsScience,Ding2016ACSAppl.Mater.Interfaces,Choudhary2020arXiv,boland2022computational}. However, so far, the existing approaches aid growth and calculation decisions for a specific pair of materials rather than allowing for a comprehensive query to yield fast, approximate predictions over a wide range of possible interfaces. 

In this paper, we introduce InterMatch, which uses information readily available from preexisting materials databases such as the Materials Project and 2DMatPedia databases to predict charge transfer\cite{Milnes1972}, strain\cite{Landau1986}, stability\cite{Muller1994PAC}, and optimal superlattice\cite{Kaxiras2003Cambridge} of an atomic interface. Using these predictions, InterMatch can narrow the candidate pool from $C>\mathcal{O}(10^6)$ to $\mathcal O(10)$ that can then be investigated in greater detail using MINT or supercell DFT (see Fig. \ref{Fig:fig1}(a)). We first illustrate how two branches of InterMatch predict the charge transfer and optimal superlattice after querying the entries of the Materials Project for each of the constituents of the interface. We then benchmark InterMatch predictions for the charge transfer against experimental measurements and supercell DFT predictions. We then employ InterMatch's branches to address two bottleneck problems obstructing design of interfaces towards the goal of discovering new physics: the problem of doping transition metal dichalcogenides and the problem of predicting stable interface structure, applied to the graphene/$\alpha$-RuCl$_3$ system.
We comment on many other classes of interfaces that can be optimized using InterMatch. 

Starting with \textit{ab initio} materials data, InterMatch performs high-throughput screening of possible heterostructures via pairwise calculation of desired interface properties including charge transfer $\Delta n$, strain tensor $\tilde{\varepsilon}$, optimized superlattice vectors $\vec{v_{1}},$ $\vec{v_{2}}$, and number of atoms $N$. Once InterMatch identifies a promising pool of candidate combinations, one can make a more in-depth analysis of the smaller pool using MINT or supercell DFT (See Fig \ref{Fig:fig1}(a)). The InterMatch algorithm has two branches to predict two key electronic and mechanical characteristics of candidate interfaces: charge transfer and optimized superlattice structure (See Fig.~\ref{Fig:fig1}).  
One branch is devoted to calculating charge transfer (Fig. \ref{Fig:fig1}(c)-(d)), and the other is devoted to optimizing supercell structure by minimizing the number of atoms and elastic energy (Fig. \ref{Fig:fig1}(e)-(f)).
For the first branch that estimates the direction and magnitude of charge transfer, we use a simple model to describe the Fermi level shifts occurring in each material when they are brought together in proximity\cite{Ruan1987JournalofAppliedPhysics}. 
Fig. \ref{Fig:fig1}(c) shows how the Fermi level shifts are determined: systems 1 and 2 are designated as ``donor" or ``acceptor" based on their relative Fermi levels $E_{F}^{1}$ and $E_{F}^{2}$, and the difference of integrals over $g_1(E)$ and $g_2(E)$
\begin{equation}
    \int_{E_{F}^{'}}^{E_{F}^{1}} dE \; g_{1}(E) = \int_{E_{F}^{2}}^{E_{F}^{'}} dE \; g_{2}(E)
\end{equation}
is minimized to determine the equilibrium Fermi level $E_{F}^{'}$. 
We take interaction between the two systems at the interface into account in the estimation of the charge transfer $\Delta n$ using a simple capacitor model\cite{Lowell1980Adv.Phys.}. Specifically, we model the interface as a parallel plate capacitor with 
 the separation $d$ given by the sum of the largest van der Waals radii of the species in each system 1 and 2 (See Fig.\ \ref{Fig:fig1}(d)). The charge transfer depends on the equilibrium Fermi level $E_F'$ and the distance $d$ as 
$e\Delta n=\varepsilon_{0} E_{F}^{'}/d$.

\begin{figure*}[t]
\begin{centering}
\includegraphics[width=.9\textwidth]{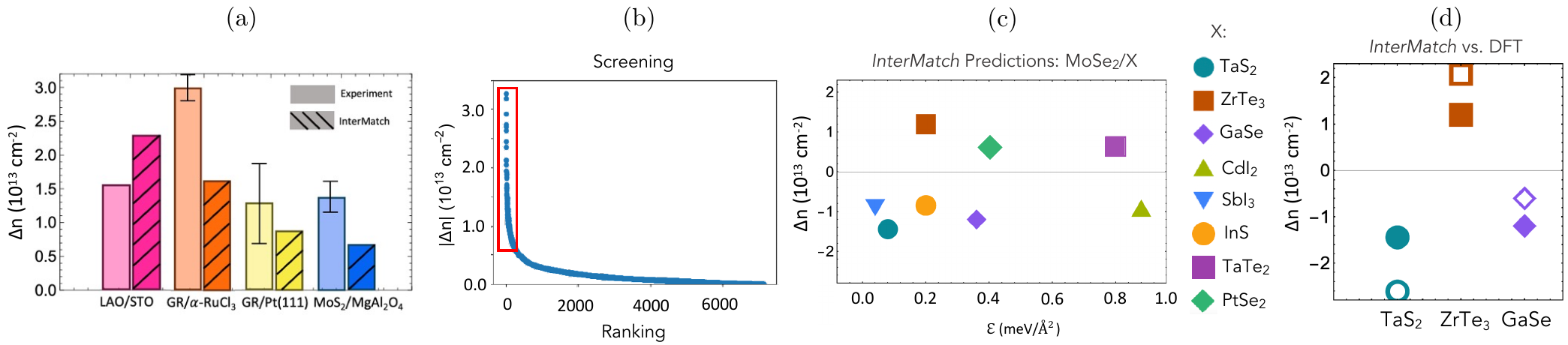}
\end{centering}
\caption{Benchmarking InterMatch. (a) Comparison of charge transfer predicted by InterMatch with measured experimental values for interfaces in Refs\cite{Annadi2013Nat.Comm,Wang2020NanoLett.,Sutter2009PhysRevB,Zheng2021Appl.Phys.Lett.}. 
(b) InterMatch screening of over 10,000 2D materials in heterostructure with monolayer MoSe$_2$, ranked in descending order of charge transfer $\abs{\Delta n}$. (c) Substrate selection based on InterMatch screening results from red box in (a) according to $\Delta n$, elastic energy $\mathcal{E}$, and energy above-hull. (d) Comparison of InterMatch predictions for $\Delta n$ (solid symbols) with supercell DFT calculations (open symbols).}
\label{Fig:fig2}
\end{figure*}

The second branch of the InterMatch algorithm sketched in Fig. \ref{Fig:fig1}(e-f) constructs optimal supercells from a pair of queried systems by calculating strain and elastic energy at their interface over a series of supercell configurations. Given the lattice vectors of system 1, $\vec{a}_1$, $\vec{b}_1$, and those of system 2, $\vec{a}_2$, $\vec{b}_2$, the algorithm searches for pairs of near-equivalent superlattice vectors
$\{(\vec{u}_1,\vec{v}_1),(\vec{u}_2,\vec{v}_2)\}$: 
\begin{equation}
\begin{aligned}
\vec{u}_i &= \mathbf{M}^{i}_{11}\vec{a}_i + \mathbf{M}^{i}_{12}\vec{b}_i\\
\vec{v}_i &= \mathbf{M}^{i}_{21}\vec{a}_i + \mathbf{M}^{i}_{22}\vec{b}_i
\end{aligned}
\label{eq:uv}
\end{equation}

where $\mathbf{M}^{i}$ is a  $2\times 2$ matrix of integer coefficients for the system $i$ and the ``near-equivalence'' is defined by 
\begin{equation}
\mathbf{M}^1=(\tilde{\varepsilon}^{2}+1)\mathcal{R}_{\theta}\mathbf{M}^2.
\label{eq:M1M2}
\end{equation}
Here $\mathcal{R}_{\theta}$ is an in-plane rotation matrix  by angle $\theta$ and $\tilde{\varepsilon}^{2}$
is the strain tensor resulting from straining the Bravais lattice of system 2 to match that of system 1.

We choose system 2 to be the material with the smallest elements of the stiffness tensor $\mathbf{C}$ (queried from the Materials Project) in the strain direction. InterMatch then computes the elastic energy $\mathcal{E}=\frac{1}{2} C_{ijkl}\varepsilon_{ij}\varepsilon_{kl}$ for the superlattice candidate according to classical elastic plate theory\cite{Landau1986}. The optimal supercell is determined by simultaneously minimizing the elastic energy $\mathcal{E}$ and the cell area $\abs{\vec{u}_i\cross \vec{v}_i}$. The use of elastic energy goes beyond previous approaches for finding the superlattice\cite{Stradi2017J.Phys.:Condens.Matter,Lazic2015ComputerPhysicsCommunications} which only consider geometric strain.

\begin{figure*}[t]
\begin{centering}
\includegraphics[width=.9\textwidth]{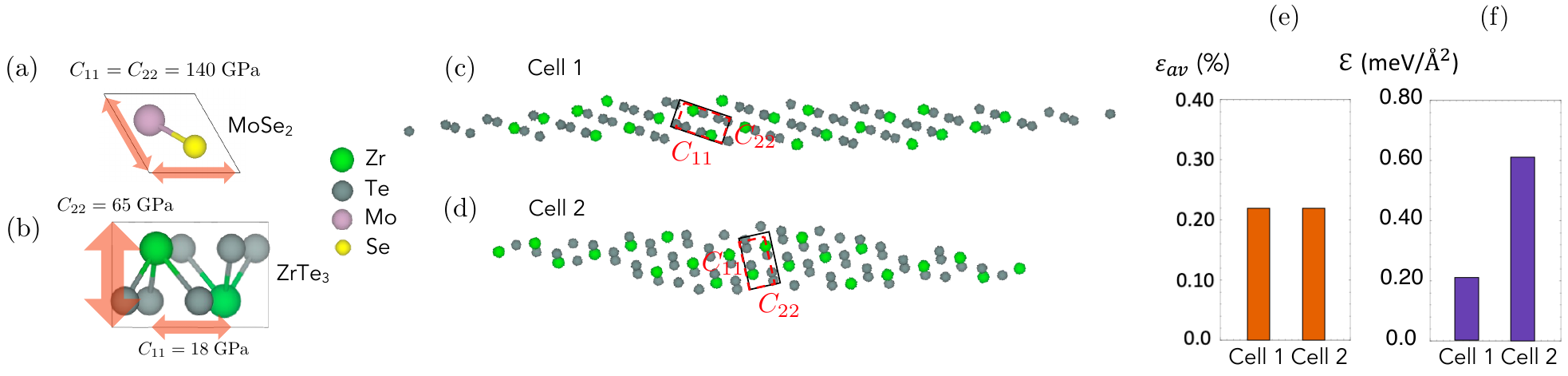}
\end{centering}
\caption{Superlattice structure prediction optimizing elastic energies. (a)-(b) (Top view)
Diagonal stiffness tensor components $C_{11}$ and $C_{22}$ of primitive MoSe$_2$ and ZrTe$_3$ unit cells in Voigt notation. (c)-(d) (Top view) ZrTe$_3$ layer of two candidate MoSe$_2$/ZrTe$_3$ supercells with the same number of atoms and average strain $\varepsilon_{av}^{\textnormal{ZrTe}_{3}}$. The solid black boxes denote the strained ZrTe$_3$ unit cells and the dashed red boxes are the original unstrained primitive cells. (e) Average strain values $\varepsilon_{av}$ of the MoSe$_2$/ZrTe$_3$ interfaces in Cells 1 and 2. (f) Elastic energies $\mathcal{E}$ of the interfaces. 
}
\label{Fig:fig3}
\end{figure*}

Now we demonstrate how elastic energy considerations can make a difference in optimization of the superlattice. 
Consider the MoSe$_2$/ZrTe$_3$ interface. 
The primitive unit cells of MoSe$_2$ and ZrTe$_3$ are shown in Fig. \ref{Fig:fig3} (a) and (b) respectively, along with the diagonal components of their stiffness tensors $C_{ii}$ in Voigt notation. 
The anisotropy of the ZrTe$_3$ stiffness tensor is such that the energetic cost to deforming ZrTe$_3$ along the direction of $C_{22}$ far exceeds the cost of an equivalent deformation along the direction of $C_{11}$. 
 Cells 1 and 2 in Fig. \ref{Fig:fig3} (c) and (d) result from an InterMatch search for low-area, low-strain MoSe$_2$/ZrTe$_3$ supercells.
The two cells are identical in number of atoms and geometric strain $\varepsilon_{av}$ (shown in Fig. \ref{Fig:fig3} (e)), however, cell 1 is favored energetically due to the different strains required to make each supercell commensurate with MoSe$_2$ (Fig. \ref{Fig:fig3} (f)).

We now benchmark charge transfer predictions by InterMatch against experimentally measured charge transfer in known interfaces.   Fig.\ \ref{Fig:fig2} (a) shows a comparison of InterMatch predictions of charge transfer with experimentally obtained values for several interfaces: LaAlO$_3$/SrTiO$_3$(1\,1\,0)\cite{Annadi2013Nat.Comm}, GR/$\alpha$-RuCl$_3$\cite{Wang2020NanoLett.}, GR/Pt$(1\,1\,1)$\cite{Sutter2009PhysRevB}, and MoS$_2$/MgAl$_2$O$_4$\cite{Zheng2021Appl.Phys.Lett.}.
The magnitudes of the $\Delta n$ predicted with InterMatch are at the same order of magnitude as the measured values, especially given experimental error bars, with the exception of GR/$\alpha$-RuCl$_3$. However, spin-orbit coupling effects (absent from our calculations) are known to affect the band structure of $\alpha$-RuCl$_3$\cite{PhysRevB.91.241110}, altering the band alignment with GR and the resulting charge transfer.

Next we turn to the application of charge transfer prediction to the problem of doping transition metal dichalcogenides (TMDs). TMDs have emerged as an exciting van der Waals material platform at the intersection of semiconductor physics and strong correlation physics\cite{Manzeli2017natrevmats}. Due to the spin-valley locking Ising spin-orbit coupling, an exotic $p$-wave superconducting state was proposed for hole-doped TMDs\cite{Hsu2017ncomms}. Recent developments in TMD moir\'e systems have further extended the phase space of possibilities. However, a major bottleneck against testing these proposals is the difficulty of establishing a good contact. Empirically, it has been established that doping the contact area can significantly improve the contact resistance\cite{Zheng2021CellReportsPhysicalScience}. However, gate-based doping does not scale well. While successful modulation doping using work function difference was established in graphene/$\alpha$-RuCl$_3$ heterostructures\cite{Wang2020NanoLett.}, it is desirable to perform an exhaustive search of interface possibilities.  

We seek 2D substrates for controlling carrier concentration in MoSe$_2$. We screen all entries of the \href{http://www.2dmatpedia.org/}{2DMatPedia database} and 3000 entries from \href{https://materialsproject.org/}{the Materials Project} for 
stable 2D materials composed of elements making up the majority of commercially available semiconductors, semimetals, and metals. 
We use InterMatch to down-sample from 10,000 candidate 2D substrates based on the magnitude of the predicted charge transfer $\abs{\Delta n}$ to MoSe$_2$ in the desired range $\gtrsim \mathcal{O}(10^{13})$ cm$^{-2}$ (Fig. \ref{Fig:fig2} (b)). We then select from these the compounds with maximum $\abs{\Delta n}$, minimal strain, and minimal above-hull energy (Fig. \ref{Fig:fig2} (c)). Finally, we choose a small subset of the top interfaces (2H-TaS$_2$, $\beta$-GaSe, and ZrTe$_3$, in the case of our example) 
and benchmark InterMatch predictions against supercell DFT calculations using the 
optimized supercells generated by InterMatch. Fig. \ref{Fig:fig2} (d) shows a comparison of InterMatch predictions with the results from DFT for the top three interfaces (for computational details, see the Supplemental Material). 
The magnitude of the $\Delta n$ prediction from the two approaches are within $10^{13}$ cm$^{-2}$. Moreover, both approaches find consistent relative magnitude of charge transfer.
Given the high-throughput nature of InterMatch, these agreements encourage using InterMatch as the first pass in searches for optimal heterostructures.

\begin{figure*}[t]
\begin{centering}
\includegraphics[width=.7\textwidth]{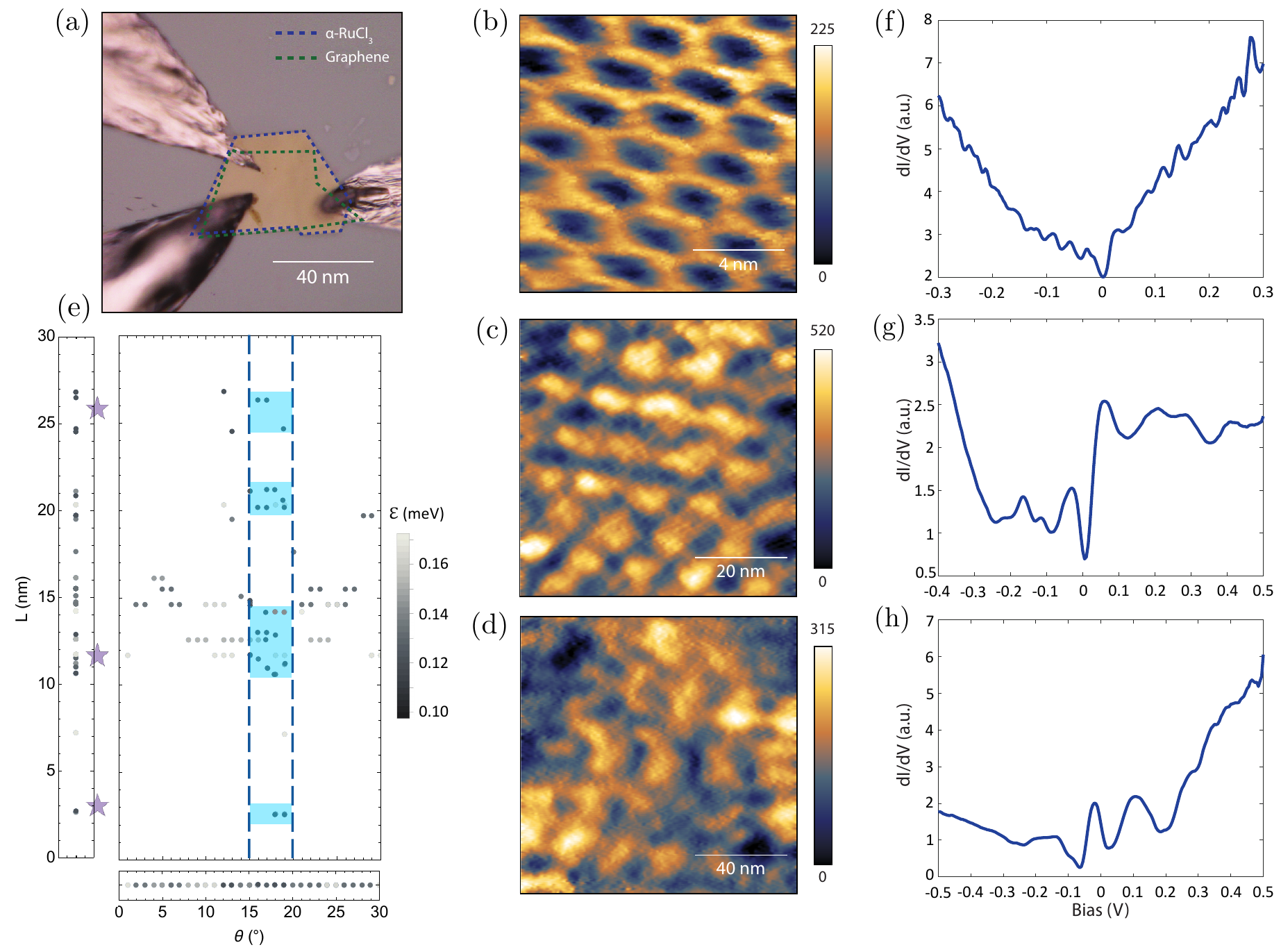}
\end{centering}
\caption{(a) Optical image of the measured device of GR/$\alpha$-RuCl$_3$ contacted with bismuth indium tin for the STM measurements. The blue and green dashed lines show the boundary of $\alpha$-RuCl$_3$ and graphene, respectively. (b)-(d) STM topographic images (in pm) of GR/$\alpha$-RuCl$_3$ on 2.7 nm (set points of -100 mV and -100 pA), 11.7 nm and 25.7nm (set points of -1 V and -50 pA) moir\'e patterns due to atomic reconstruction. (e) InterMatch predictions for low-energy GR/$\alpha$-RuCl$_3$ superlattice configurations as a function of period $L$, twist angle $\theta$, and elastic energy $\mathcal{E}$. Left panel is a projection onto the $L$-axis, bottom panel is a projection onto the $\theta$-axis. Dashed box indicates interval of $\theta$ containing largest range of stable superlattice periods, shaded blue boxes indicate regions of likely superlattice configurations. Purple stars denote the periodicities extracted from the experiment.  (f)-(h) dI/dV measurements corresponding to the three moir\'e patterns in (b)-(d) showing strong resonances dependent on moire wavelengths. 
}
\label{Fig:newfig4}
\end{figure*}

As an example of the power of Intermatch to understand superlattice structure, we consider the graphene/$\alpha$-RuCl$_{3}$ heterostructure (GR/$\alpha$-RuCl$_{3}$). This system has attracted great interest due to the presence of strong modulation doping\cite{Wang2020NanoLett.} and enhancement of $\alpha$-RuCl$_{3}$'s proximity to the Kitaev spin liquid phase. However, relatively little attention has been paid to the atomic scale structure of the heterostructure and the possible influence on electronic properties. In order to study this experimentally, we used scanning tunneling microscopy (STM) to investigate the properties of GR/$\alpha$-RuCl$_{3}$ heterostructures created by mechanical exfoliation and colamination, as shown in Fig.\ \ref{Fig:newfig4} (a). The angle between the $\alpha$-RuCl$_{3}$ substrate and graphene was not intentionally controlled. Shown in Fig.\ \ref{Fig:newfig4} (b)-(d) are a set of STM topographs taken at various locations of the GR/$\alpha$-RuCl$_{3}$ heterostructure. The locations are within a few microns of each other on the sample shown in Fig.\ \ref{Fig:newfig4} (a). Intriguingly, all three of the regions show   
moir\'e patterns with large wavelengths - 2.7 nm in Fig.\ \ref{Fig:newfig4} (b), 11.7 nm in Fig.\ \ref{Fig:newfig4} (c) and 25.7 nm in Fig.\ \ref{Fig:newfig4} (d). All three of these wavelengths are much larger than the wavelength set by the difference in lattice constants.

Using InterMatch, we perform a comprehensive mapping of the space of superlattice configurations spanned by $(\theta,L,\mathcal{E})$ where $\theta$ is the twist angle, $L$ is the moir\'e period, and $\mathcal{E}$ is the elastic energy of the interface. The resulting spectrum of low-energy superlattice configurations for $0^{\circ} \leq \theta \leq 30^{\circ}$ and $0\; \textnormal{nm} \leq L \leq 30\; \textnormal{nm}$ is shown in Fig.\ \ref{Fig:newfig4} (e). We identify four prominent moir\'e length scales (blue boxes in Fig.\ \ref{Fig:newfig4} (e)) occurring within a $5^{\circ}$ range between $15^{\circ}$-$20^{\circ}$. Three of the four length scales coincide with those observed in STM at $L=2.7,\;11.7,\;25.7$ nm, shown in Fig.\ \ref{Fig:newfig4} (b)-(d). Correctly identifying energetically favorable GR/$\alpha$-RuCl$_{3}$ superlattices over a narrow range of twist angles showcases InterMatch's capability to predict interfacial structure of complex (e.g extremely lattice-mismatched) systems.

The presence of an atomic reconstruction at the interface of GR/$\alpha$-RuCl$_{3}$ can have dramatic consequences for the spectroscopic properties of the material. Shown in Fig.\ \ref{Fig:newfig4} (f)-(h) are scanning tunneling spectra averaged over the regions shown in Fig.\ \ref{Fig:newfig4} (b)-(d). These spectra show dramatic differences from the simple expectation for a doped Dirac spectrum as might be expected from charge transfer alone. Instead, we observe strong resonances in all three regions, with the spacing between resonances following the expectation from Landau levels on a Dirac spectrum. Previously, such spectra have been observed when graphene has a periodic buckling\cite{mao2020evidence}, where it was ascribed to periodic strain in the material. In our case, apart from the strain associated with the moir\'e lattice\cite{shabani2021deep}, we expect that there will also be strong periodic variations in the doping\cite{rizzo2022nanometer} that contribute to the formation of resonances.

In summary, we introduce and demonstrate InterMatch, a high-throughput computational framework and database for predicting charge transfer, strain, and superlattice of an interface between two arbitrary materials. Charge transfer allows heterostructure-based modulation doping\cite{Wang2020NanoLett.}, which can guide device fabrication and contact design
\cite{Zheng2021CellReportsPhysicalScience}.  Efficiently determining the smallest energetically favorable commensurate supercells from a wide variety of interface configurations is crucial for accelerating \textit{ab initio} studies. We showcase the use of InterMatch by identifying high-charge transfer substrates for doping TMDs, and by predicting equilibrium moir\'e superlattice configurations for the lattice-mismatched GR/$\alpha$-RuCl$_3$ interface that are validated by STM measurements. \anp{The presence of such long-wavelength superlattice modulations at van der Waals interfaces present new opportunities to tailor bandstructure using materials that do not have a close match in lattice constants.} The evolving interface database provides open access to InterMatch results which we hope will help guide future exploration of interfacial systems.

To broadly benefit the community, we made the InterMatch code openly accessible at https://doi.org/10.5281/zenodo.6823973\cite{eg5872022}. Moreover, we tabulate InterMatch results in an open-access ``interface database" directly integrated with the Materials Project via the MPContribs platform. At the time of writing, the database contains $\sim 200,000$ interfaces (and counting) in simple JavaScript Object Notation (JSON) that are queryable and sortable according to the chemical composition of either constituent system, charge transfer, strain, and optimized supercell size. In addition, we generate crystallographic information files (CIF) of interface supercells with InterMatch which may be readily accessed from the database and used as inputs for DFT or other first principles studies.

\begin{acknowledgements}
The authors thank Kin Fai Mak, Jie Shan, Stephen Carr, Patrick Huck, Matthew Horton, Jason Munro, and Vidya Madhavan for helpful discussions.
EAK and JH were supported by MURI grant FA9550-21-1-0429.  
EG was supported by the Cornell Center for Materials Research with funding from the NSF MRSEC program (DMR-1719875). SBT was supported by the Department of Energy Computational Science Graduate Fellowship under grant DE-FG02-97ER25308.
The computation was done using the high powered computing cluster WALLE2 that was established through the support of  Gordon and Betty Moore Foundation’s EPiQS Initiative, Grant GBMF10436 to EAK and  the New Frontier Grant from Cornell University’s College of Arts and Sciences and hosted and maintained by Cornell Center for Advanced Computing. \anp{STM experiments were supported by NSF DMR-2004691 (ANP) and AFOSR via grant FA9550-21-1-0378 (SS, ES). Sample synthesis for STM measurements was supported by the NSF MRSEC program through Columbia in the Center for Precision-Assembled Quantum Materials (PAQM), grant number DMR-2011738.}  
\end{acknowledgements}

\clearpage 
\newpage 
\begin{widetext}
\beginsupplement
\section*{Supplemental Material}
In Section \ref{SMsubstrates} we present additional InterMatch results identifying optimal substrate candidates for forming high-charge-transfer interfaces with group-VI transition metal dichalcogenides and putative spin liquid materials $\alpha$-RuCl$_{3}$ and TbInO$_{3}$. In Section \ref{SMcode} we provide details of the InterMatch code. In Section \ref{SMcomputational} we provide computational details of the \textit{ab initio} density-functional theory (DFT) calculations of MoSe$_2$ supercells in Fig. 2(d) of the main text.
\section{Charge transfer substrates for WT\lowercase{e}$_2$, WS\lowercase{e}$_2$, and $\alpha$-R\lowercase{u}C\lowercase{l}$_{3}$} \label{SMsubstrates}
We further demonstrate InterMatch by applying it to two types of materials: group-VI TMDs, and putative spin liquid $\alpha$-RuCl$_{3}$\cite{Plumb2014Phys.Rev.B}. TMDs' unique combination of properties makes them highly attractive for nanoelectronics applications and fundamental studies of novel physical phenomena\cite{Manzeli2017natrevmats}. However, the realization of many such high-performance devices and exotic phases is limited by the availability of systems with high carrier mobility and low contact resistances between metal contacts and the semiconductor. Quantum spin liquids (QSLs) are interacting quantum systems in which spins do not order at low temperatures, and have been theorized to offer insights into high-temperature superconductivity upon doping. $\alpha$-RuCl$_{3}$, for example, has been intensively discussed as a possible candidate for Kitaev physics; however, it orders antiferromagnetically at low temperatures due to the presence of additional magnetic couplings extending beyond the pure Kitaev interaction. Doping $\alpha$-RuCl$_{3}$ with charge carriers has been predicted to enhance Kitaev interactions and push $\alpha$-RuCl$_{3}$ closer to the spin liquid phase. We use InterMatch to identify stable, high-charge-transfer interfaces for electron- and hole-doping the TMDs WTe$_2$, WSe$_2$, and MoSe$_2$, and the putative QSL $\alpha$-RuCl$_{3}$. Fig. \ref{Fig:fig5} shows sample InterMatch results of substrate candidates for interfaces with WTe$_2$, WSe$_2$, and $\alpha$-RuCl$_{3}$, highlighting those that minimize elastic energy $\mathcal{E}$ and maximize charge transfer $\Delta n$. In Fig. \ref{Fig:fig5} (a)-(b), we survey all $\sim 70,000$ oxides from the Materials Project. Oxides have a wide range of charge neutrality levels and therefore constitute a powerful addition to electrostatic gating or chemical doping for controlling carrier concentration in heterostructures. In Fig. \ref{Fig:fig5} (c) we survey all entries in the 2DMatpedia database, and determine two-dimensional (2D) substrate candidates to maximally dope $\alpha$-RuCl$_{3}$.
\begin{figure}[h]
\begin{centering}
\includegraphics[width=\textwidth]{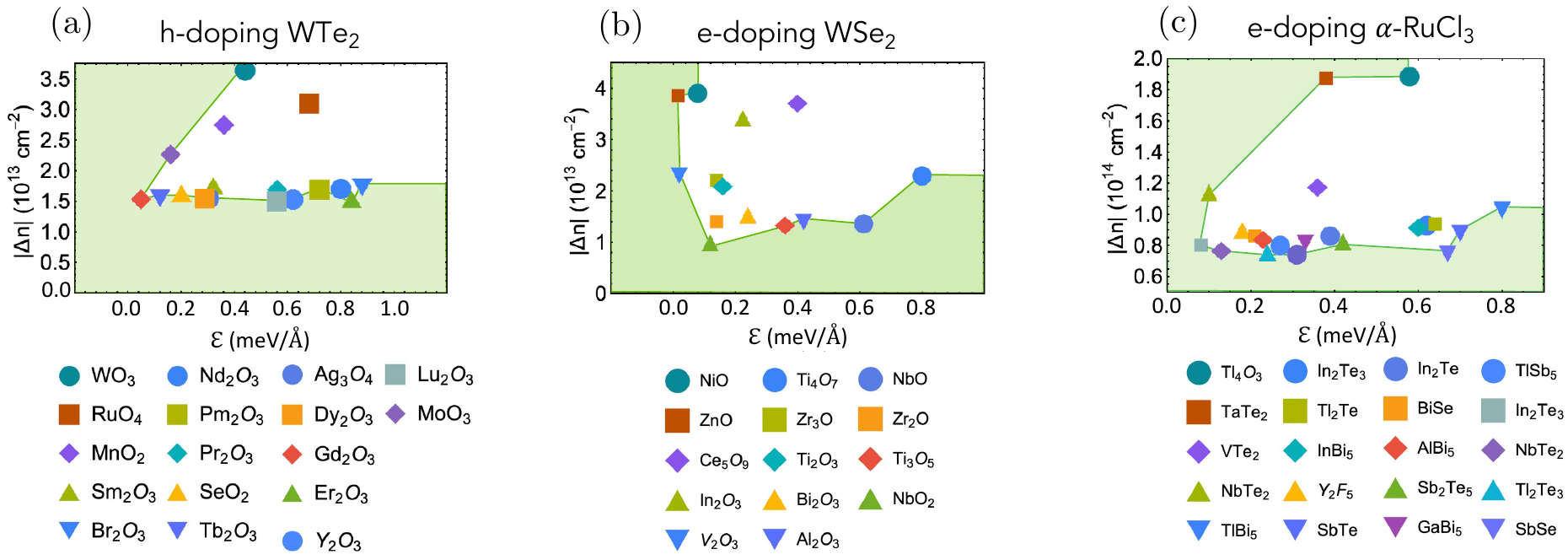}
\end{centering}
\caption{Sample InterMatch results for different charge transfer interfaces. The solid green line indicates the pareto-optimal frontier of substrate candidates that minimize elastic energy $\mathcal{E}$ and maximize charge transfer $\abs{\Delta n}$. Calculated quantities in (a)-(b) use \textit{ab initio} data from the Materials Project as inputs, while those in (c) use data from the 2DMatpedia database.
}
\label{Fig:fig5}
\end{figure}
\section{InterMatch Code} \label{SMcode}
The InterMatch code is written in Python 3.7 and makes extensive use of pymatgen\cite{Ong2013ComputationalMaterialsScience}, an open-source Python package of the Materials Project, for the manipulation and analysis of various structures of interest. The code is continuously being developed, and the latest version can be obtained at https://doi.org/10.5281/zenodo.6823973\cite{eg5872022}. We aim to provide an efficient scheme for computing interface properties capable of screening a significant fraction of combinations of existing Materials Project structure entries, returning the results in real-time (typical run time for the calculation of a single interface is $10 \pm 5$ seconds on a 1.7 GHz Intel Core i5 processor at 1333 MHz using 4 GB of RAM, running macOS Sierra 10.12.6).

\section{Computational Details} \label{SMcomputational}
All \textit{ab initio} DFT calculations were carried out within the total-energy plane wave density-functional pseudopotential approach, using Perdew-Burke-Ernzerhof generalized gradient approximation functionals\cite{Perdew1996PRL} and optimized norm-conserving Vanderbilt pseudopotentials in the SG15 family\cite{Schlipf2015Comp.Phys.Comms}. Plane wave basis sets with energy cutoffs of 30 hartree were used to expand the electronic wave functions. We used fully periodic boundary conditions and a $8 \times 8 \times 1$ $k$-point mesh to sample the Brillouin zone. Electronic minimizations were carried out using the analytically continued functional approach starting with a LCAO initial guess within the DFT$++$ formalism\cite{Freysoldt2009PRB}, as implemented in the open-source code JDFTx\cite{Sundararaman2017SoftwareX} using direct minimization via the conjugate gradients algorithm\cite{Payne1992RMP}. All unit cells were constructed to be inversion symmetric about $z=0$ with a distance of $\sim 60$ bohr between periodic images of the MoSe$_{2}$ surface, using coulomb truncation to prevent image interaction.
\end{widetext}
\bibliography{intermatchbib}

\end{document}